# On Power Control and Frequency Reuse in the Two User Cognitive Channel

Onur Ozan Koyluoglu and Hesham El Gamal

**Abstract**

This paper considers the *generalized* cognitive radio channel where the secondary user is allowed to reuse the frequency during *both* the idle and active periods of the primary user, as long as the primary rate remains the same. In this setting, the optimal power allocation policy with single-input single-output (SISO) primary and secondary channels is explored. Interestingly, the offered gain resulting from the frequency reuse during the active periods of the spectrum is shown to disappear in both the low and high signal-to-noise ratio (SNR) regimes. We then argue that this drawback in the high SNR region can be avoided by equipping both the primary and secondary transmitters with multiple antennas. Finally, the scenario consisting of SISO primary and multi-input multi-output (MIMO) secondary channels is investigated. Here, a simple Zero-Forcing approach is shown to significantly outperform the celebrated Decoding-Forwarding-Dirty Paper Coding strategy (especially in the high SNR regime).

## I. BACKGROUND

In the classical cognitive radio set-up, the secondary users[1] must first sense the wireless channel to determine the empty parts of the spectrum. Those users will then transmit their own messages over these *white* spaces in order to increase the overall spectral efficiency. In other words, the cognitive radios can only transmit over those particular frequency bands (or time intervals) in which the licensed (primary) users are not transmitting. Recent studies, however, have introduced *the generalized cognitive radio* concept in which the secondary user can exploit the active areas in the spectrum (i.e., simultaneously transmits with the primary users) as long as certain *co-existence* constraints are satisfied [1]–[6]. In the extreme case, where the primary user is willing to accommodate the needs of the secondary user, one can easily envision cooperation strategies where the *two* users can benefit [1]. Interestingly, even in the other extreme, where the primary user is ignorant of the secondary user presence, it was argued recently that frequency reuse is possible at secondary nodes during active primary period [3], [5]. Here, we focus on the latter approach and revisit the conclusion drawn in [3], [5] under a more realistic assumption on the secondary user side information. Furthermore, we study a generalized setting that allows for frequency re-use during both the active and idle periods of the spectrum.

More precisely, our generalized cognitive radio set-up corresponds to a two state primary channel. In **state** 1, the primary transmitter is silent and the secondary user is free to use the channel in any arbitrary way. In **state** 2, the primary user is active and the secondary user is allowed to re-use the spectrum as long as the following *coexistence constraints* are satisfied [5], [6].
- The primary encoder and decoder have the same structure as in the non-cognitive scenario.
- The primary user achieves the same instantaneous rate as in the non-cognitive scenario.

This work is submitted to the IEEE Transactions on Wireless Communications on October 7, 2007; and revised on October 16, 2008.

The authors are with the Department of Electrical and Computer Engineering, The Ohio State University, Columbus, OH 43210, USA (e-mail: {koyluogo,helgamal}@ece.osu.edu). Hesham El Gamal also serves as the Director for the Wireless Intelligent Networks Center (WINC), Nile University, Cairo, Egypt.

The material in this paper was presented in part at the IEEE International Symposium on Information Theory, Nice, France in June 2007.

We acknowledge the generous funding of the National Science Foundation and Texas Instruments.

[1]We will use terms secondary users and cognitive radios interchangeably throughout the paper.

Our work attempts to quantify the potential gain of re-using the spectrum in state 2. Towards this end, we derive power allocation policies and propose cognitive transmission schemes in three distinct scenarios. The first one assumes SISO primary and secondary channels. Here, the gain offered by reusing the spectrum in state 2 with the Decoding-Forwarding-Dirty Paper Coding (DF-DPC) scheme [3], [5] is shown to disappear in both the low and high SNR regimes. The scheme is then argued to fail in MISO-MISO scenario, where both the primary and secondary transmitters are equipped with multi-antennas. In this scenario, although the limitation arising from the decoding time can be mitigated, the need to forward the primary signal limits the achievable cognitive rate. In particular, the gain resulting from the utilization of the active primary states can only be in the form of an SNR offset gain with this scheme. This observation motivates the novel Decoding-Dirty Paper Coding-Zero Forcing (D-DPC-ZF) approach, which is shown to achieve significant gains in the high SNR regime by efficiently re-using the spectrum in state 2. The scheme of [3], [5] is also argued to fail in the SISO-MIMO channel. To avoid this drawback, we propose a simple Zero Forcing (ZF) scheme capable of efficiently re-using the spectrum in state 2 of the generalized SISO-MIMO channel.

A brief remark about the difference between our work and earlier information theoretic studies [1], [3]–[6] is now in order. The first differentiating factor is the generalized cognitive radio notion which offers a realistic scenario where the cognitive user is able to exploit both the idle and busy areas of the spectrum (as opposed to earlier approaches which focuses only on the active areas). Secondly, we relax the unrealistic assumption, adopted in [3]–[6], where the secondary transmitter is assumed to have non-causal information about the primary message. We note that [1] proposes some causal schemes, where it is assumed that the primary users are willing to cooperate with the secondary users. As shown in the sequel, these two factors induce fundamental changes in the problem. For example, in the SISO-SISO scenario the time needed to decode the primary message at the cognitive transmitter is shown to limit the ability of re-using the active spectrum in the high SNR regime (a result which contradicts the spirit of the conclusions in [3]–[6]).

The rest of this work is organized as follows. In Section II, the system model and necessary notations are introduced. Section III is devoted to the SISO-SISO scenario. The analysis for the MISO-MISO channel along with the proposed D-DPC-ZF scheme is detailed in Section IV. Section V presents the ZF strategy which achieves a significant performance gain in the SISO-MIMO generalized cognitive channel. Finally, we offer some concluding remarks in Section VI. To enhance the flow of the paper, the proofs are collected in the Appendix.

## II. System Model and Notations

We consider a four-terminal network, in which the primary transmitter and receiver are Nodes 1 and 3, respectively; whereas their secondary counterparts are Nodes 2 and 4, respectively. All nodes are assumed to be **half-duplex**, and each receive antenna is impaired by an additive independent and identically distributed circularly symmetric complex Gaussian noise with zero mean and unit variance, which is denoted by $\mathcal{CN}(0,1)$. The transmitters are limited by individual long-term average power constraints: $\beta P$ and $P$ for the primary and cognitive transmitters, respectively. We further assume that the secondary nodes have perfect knowledge about the channel, i.e., the cognitive radios know all channel coefficients which are assumed to be fixed complex-valued numbers. We set the magnitude square of the channel gain from Node $i$ to Node $j$ according to the path-loss model: $|h_{ij}|^2 = d_{ij}^{-2}$, where $d_{ij}$ denotes the distance between the corresponding nodes. For multi-antenna scenarios, the channel coefficient between the $k^{th}$ antenna at node $i$ and the $m^{th}$ antenna at node $j$ will be represented by $h_{ij,km}$. We denote this in the matrix form $\mathbf{H}_{ij}$, whose $m$th row and $k$th column corresponds to $h_{ij,km}$. We denote the channel coefficients in the vector form, i.e., $\mathbf{h}_{ij}$, if only one user has multi-antennas.

We adopt the asymptotic assumption of $B \to \infty$ blocks with $N \to \infty$ channel uses per block. We remark that one can simply imagine blocks as different frequency bands on which cognitive radio is active. It is further assumed that the primary transmitter is silent, i.e., in state 1, in any particular block with probability $p$ and the cognitive user is informed *a-priori* with **only** the states of the different blocks. Mathematically, we denote the instantaneous cognitive rate during states 1 and 2 as $R_1(P_1)$ and $R_2(P_2)$, respectively. Hence, our power allocation problem can be stated as follows:

$$R^{(g)} = \max_{P_1,P_2: pP_1+(1-p)P_2 \leq P} \left\{ pR_1(P_1) + (1-p)R_2(P_2), \right\} \tag{1}$$

where $R^{(g)}$ stands for the rate achieved by any generalized cognitive scheme. Here, for some $t \in [0,1]$, we set $P_1 = \frac{P(1-t)}{p}$ and $P_2 = \frac{Pt}{(1-p)}$ to parameterize the power allocation problem using the power allocation parameter $t$ as below.

$$R^{(g)} = \max_{t \in [0,1]} \left\{ pR_1\left(\frac{P(1-t)}{p}\right) + (1-p)R_2\left(\frac{Pt}{(1-p)}\right) \right\} \tag{2}$$

We denote the optimal power allocation parameter in (2) as $t^*$ and denote the rate achieved by the classical approach as $R^{(c)}$, which corresponds to $t=0$. We say that the above cognitive rate is achievable if there exists a cognitive coding/decoding scheme which satisfies the *coexistence constraints* and allow both receivers to decode their messages with arbitrarily small error probabilities. For a given cognitive scheme, we respectively denote the multiplexing gain [7] (a.k.a. degrees of freedom) of the generalized and the classical approach as [2]

$$r^{(g)} \triangleq \lim_{P \to \infty} \frac{R^{(g)}}{\log(P)}, \text{ and } r^{(c)} \triangleq \lim_{P \to \infty} \frac{R^{(c)}}{\log(P)}, \tag{3}$$

where $P$ is the SNR at the cognitive receiver as the noise at each receiver antenna is unitary.

## III. SISO-SISO Channel

In this section, we analyze the power allocation problem of cognitive radio channels under the assumption of single-input-single-output (SISO) primary and cognitive links. The main hurdle now is to identify the optimal coding strategy which allows the secondary user to efficiently re-use the active primary period. Instead of pursuing this problem, which appears intractable at the moment, we assume that the system is in the low-interference-gain regime and the cognitive user will implement the scheme proposed in [3]–[6] during state 2, i.e., the cognitive transmitter will first decode the primary message in $N'$ channel uses (assuming $|h_{13}| \leq |h_{12}|$), and then, the cognitive transmitter will send its own message using Dirty Paper Coding (DPC) [11]. In order to maintain a fixed instantaneous rate of primary link, only a fraction of the available power will be allocated to this signal. The cognitive transmitter will use the remaining power to cooperate with the primary user in forwarding its message. The power allocation parameter during this transmit period, represented as $u$ below, should be judiciously chosen such that the cooperation benefit will *exactly* compensate for the interference caused by the secondary signal. We note that in the weak-interference regime, i.e., ($|h_{23}| \leq |h_{24}|$), this is the capacity achieving scheme under the assumption that the cognitive user is informed *a-priori* with the primary message [3], [4]. However, it turns out that the generalized cognitive radio with this scheme will reduce to the classical one in both low and high SNR regions, as shown in the following theorem which also gives the optimal power control policy for the secondary user.

*Theorem 1:* An achievable rate of the SISO cognitive link using the DF-DPC scheme can be denoted as follows:

$$\begin{aligned} R^{(g)}_{\text{DF-DPC}} &= \max_t \left\{ p \log\left(1 + \frac{|h_{24}|^2 P(1-t)}{p}\right) \right. \\ &\quad + (1-p)(1-\alpha) \log\left(1 + \frac{|h_{24}|^2 Ptu}{(1-p)(1-\alpha)}\right) \right\}, \end{aligned} \tag{4}$$

---
[2] Throughout the sequel, all logarithms are taken to base-2.

where

$$\alpha = \frac{\log\left(1 + \frac{|h_{13}|^2 \beta P}{(1-p)}\right)}{\log\left(1 + \frac{|h_{12}|^2 \beta P}{(1-p)}\right)} \quad (5)$$

$$u = 1 - \left(\frac{|h_{13}|\sqrt{\beta}\left(-\sqrt{(1-\alpha)}(1-p) \mp \sqrt{\delta}\right)}{|h_{23}|\sqrt{t}\left(1-p+|h_{13}|^2\beta P\right)}\right)^2$$

$$\delta = (1-\alpha)(1-p)^2 + P|h_{23}|^2 t\left(1-p+|h_{13}|^2\beta P\right),$$

and $u \in [0, 1]$.

Moreover,

$$\lim_{P \to 0} \frac{R^{(g)}_{\text{DF-DPC}} - R^{(c)}}{P} = 0, \quad \lim_{P \to \infty} R^{(g)}_{\text{DF-DPC}} - R^{(c)} = 0.$$

*Proof:* Please refer to Appendix I. ■

We note that, $\alpha$ represents the fraction of an active primary block during which cognitive radio listens the primary transmission and $u$ is the fraction of the power allocated to the cognitive signal during active primary states. We remark that the limiting behavior established in Theorem 1 implies that DF-DPC scheme reduces to the classical cognitive channel where the secondary transmitter is only active in the silent periods of the primary link (i.e., the optimal point for (4) approaches to $0$). The question now is whether this negative result, in the low and high SNR regimes, is an implication of using the DF-DPC scheme by the cognitive user. If true, then one should be motivated to seek a better cognitive transmission scheme.

Remarkably, the following argument shows that our negative result is a fundamental property of the low SNR regime. In other words, there is no cognitive transmission scheme which satisfies the coexistence constraints and achieves a higher low SNR gain than the classical approach, if the probability of primary silence state is non-zero (i.e., $p > 0$). Our primary user is totally ignorant of the cognitive user presence, and hence, is not willing to cooperate. Then, removing the primary user from the system can only increase the achievable rate of the cognitive user. In this idealistic scenario, it is clear that the cognitive transmitter does not need to use state $2$ to transmit its message in the low SNR region (as it is power limited). Mathematically, assuming $p \neq 0$, a classical cognitive radio will only use the first state and achieve a rate of $R^{(c)} = pR_1\left(\frac{P}{p}\right)$, while a generalized one considers both states for its transmission and, by implementing power allocation, can achieve a rate of $R^{(g)}(t) = pR_1\left(\frac{P(1-t)}{p}\right) + (1-p)R_2\left(\frac{Pt}{(1-p)}\right)$ for any $t \in [0, 1]$. Here, $R_1$ and $R_2$ are the same functions: For example, in the single-antenna setting, $R_1(P) = R_2(P) = \log(1 + |h_{24}|^2 P)$, as we have removed the primary user, and hence

$$\lim_{P \to 0} \frac{R^{(g)} - R^{(c)}}{P} = 0,$$

where we used the fact that $\lim_{P \to 0} \frac{\log(1+hP)}{\log(e)P} = h$.

Now, we proceed to the high SNR regime. The culprits behind the negative result of the DF-DPC scheme in the high SNR regime are: 1) The decoding time required by the cognitive transmitter to figure out the primary message, which dominates the whole block asymptotically (i.e., $\alpha \to 1$ as $P \to \infty$), and 2) The fraction of power allocated to the cognitive signal after decoding the primary message, which diminishes as power gets large (i.e., $u \to 0$ as $P \to \infty$). To probe further, we first analyze the non-causal (genie-aided) scenario [4], [6], for which we assume that the primary message is non-causally given to the cognitive transmitter at the beginning of each active primary block. Note that, the cognitive transmitter does not need to decode the primary message as the message is already given. We note the following observation for this over-realistic scenario.

*Corollary 1:*
$$\lim_{P \to \infty} R_{\text{F-DPC}}^{(\text{g,nc})} - R^{(\text{c})} = G,$$

where $R_{\text{F-DPC}}^{(\text{g,nc})}$ is the achievable rate of the generalized non-causal cognitive link using the Forwarding-Dirty Paper Coding (F-DPC) scheme, and $G$ is a constant depending on the channel parameters.

*Proof:* Please refer to Appendix II. ∎

Hence, in the non-causal case, cognitive radios can utilize the active primary periods to achieve a better SNR offset compared to the classical approach. However, as shown in Theorem 1, once we impose the causality constraint and decode the primary message, this rate gain obtained from the active primary periods disappears in the high power regime.

The previous claims are validated numerically in Fig. 2, which uses the linearized channel model of Fig. 1 (with a path loss exponent of 2). This figure shows the gain offered by the generalized cognitive radio, as compared with the classical cognitive radio, to be significant only in the medium SNR regime. Note that, the gain is even more diminishing as the distance between the transmitters in the linearized channel model increases, once we keep the distance between the cognitive users fixed. Moreover, the achievable rate for the non-causal generalized approach, i.e., $R_{\text{F-DPC}}^{(\text{g,nc})}$, is shown to have an SNR offset gain, which disappears for the causal case in the high SNR regime. While this result pertains only to the scheme proposed in [4], [6] for the causal scenario, we remark that this approach is optimal (for the weak-interference regime, i.e., $|h_{23}| \leq |h_{24}|$) among the class of SISO cognitive schemes that require decoding of the primary message at the secondary transmitter. We further note that there is no known SISO cognitive scheme which allows the frequency reuse of the active primary channel while satisfying the co-existence constraints without requiring the cognitive transmitter to decode the primary message. In fact, using the result of [8], one can easily establish the **non-existence** of any cognitive transmission scheme which achieves a non-zero DoF gain for the generalized cognitive user compared to the classical one. The argument is as follows. [8] shows that the multiplexing gain for the sum-rate of the two-user channel with full-duplex nodes together with transmitter and receiver cooperation is 1. This is an upper bound for the sum-rate multiplexing gain during the active primary states. As we need to satisfy the coexistence constraints, the multiplexing gain of the cognitive link can only be 0 during the active primary states. In other words,

$$\lim_{P \to \infty} \frac{R^{(\text{g})} - R^{(\text{c})}}{\log(P)} = 0,$$

for any generalized cognitive radio scheme. We remark that this conclusion even holds for any non-causal generalized cognitive radio scheme, as the multiplexing gain for the sum-rate in the non-causal case is 1 [9]. Therefore, the best one can hope for is a cognitive rate which grows sub-logarithmically with the SNR during the active primary state for the strong interference regime, i.e., $|h_{23}| > |h_{24}|$. Overall, our results in this section can be used to argue that re-using the active spectrum areas in our SISO-SISO set-up only offers a limited performance gain which disappears in the low and high SNR regimes.

## IV. MISO-MISO CHANNEL

In an attempt to avoid the inefficiency of SISO-SISO cognitive links on re-using the active spectrum in the high SNR regime, we now start equipping the transmitters/receivers with multiple antennas. Our first step is to consider a cognitive user with multiple transmit antennas. Unfortunately, this modification will not solve our problem. To see this, one needs to recall that the main source for inefficiency in the SISO-SISO scheme was the required decoding time at the *cognitive transmitter*. One may tempted to eliminate the decoding stage now since having multiple antennas allows the cognitive user to cancel its own interference at the primary receiver without needing the side information about the primary message. This way, however, the primary transmission will cause excessive interference at the cognitive receiver, and consequently, the benefit resulting from re-using the spectrum in state 2 can only be in the form of a fixed gain in the SNR offset in the high SNR regime. This discussion motivates the MISO-MISO scenario where both the primary and cognitive transmitters have multiple antennas.

In particular, we consider a system in which the transmitters are equipped with $M$ antennas. We assume that the primary transmitter is not informed *a-priori* about the channel state information (CSI) except for the transmission rate, $R_p$ which can be received correctly at primary receiver (i.e., the primary channel capacity in absence of the cognitive user). We further assume that the primary user employs codebook drawn randomly from an i.i.d. Gaussian codebook. Therefore, the covariance matrix of the primary signal $\mathbf{x}_1$ is $\frac{\beta P}{(1-p)M}\mathbf{I}_M$ which satisfies the average power constraint of the primary transmitter.

In this setting, we analyze a cognitive transmitter which implements the D-DPC-ZF scheme. As in the previous section, it will first decode the primary message in $N'$ channel uses of the active primary period. Then, it will start to transmit its own dirty paper coded signal. This time, however, the secondary transmitter use ZF to cancel the interference at the primary receiver by appropriately choosing a transmit beamforming vector. The following result characterizes the optimal power control policy with the proposed scheme.

*Theorem 2:* An achievable rate of the MISO-MISO cognitive channel using the D-DPC-ZF scheme can be denoted as follows:

$$R_{\text{D-DPC-ZF}}^{(g)} = \max_t \left\{ p \log \left( 1 + \frac{P(1-t)}{p} \sum_{i=1}^{M} |h_{24,i}|^2 \right) + (1-p)R_2 \right\},$$

with

$$R_2 = \begin{cases} (1-\alpha) \log \left( 1 + \frac{Pt}{(1-\alpha)(1-p)} \sum_{i=1}^{M-1} |\tilde{h}_{24,i}|^2 \right), & \text{if } \alpha \leq 1 \\ 0, & \text{if } \alpha > 1 \end{cases}$$

where

$$\alpha = \frac{\log \left( 1 + \frac{\beta P}{(1-p)M} \mathbf{h}_{13} \mathbf{h}_{13}^H \right)}{\log \det \left[ \mathbf{I}_M + \frac{\beta P}{(1-p)M} \mathbf{H}_{12} \mathbf{H}_{12}^H \right]},$$

$\tilde{\mathbf{h}}_{24} \triangleq \mathbf{h}_{24} \mathbf{Q}_t$, and $\mathbf{Q}_t$ is a projection matrix at the cognitive transmitter, orthonormal columns of which are orthogonal to $\mathbf{h}_{23}$.

Furthermore, the multiplexing gain with this scheme is given by $r_{\text{D-DPC-ZF}}^{(g)} = p + (1-p)(1-\frac{1}{M})$, while the multiplexing gain of the classical approach is $r^{(c)} = p$.

*Proof:* Please refer to Appendix III. ∎

Here, the gain of this scheme is evident in the expression of the listening interval, i.e., $\alpha$. Under the mild assumption that the channel gain matrix $\mathbf{H}_{12}$ is not rank deficient, the listening rate scales as $M \log(P)$ in the high SNR regime. Thus, $\alpha \to \frac{1}{M}$ as $P \to \infty$. For instance, adding one extra antenna at each transmitter in the system will give $\alpha \approx 0.5$ in the high SNR regime. Therefore, contrary to the previous section, the remaining fraction of the block (in state 2) can be exploited by cognitive user, resulting in a cognitive rate that scales as $0.5 \log(P)$ in the high SNR regime during state 2. These observations are validated by numerical results in Fig. 3 with Rayleigh fading channel gains, where we set the variances of the channel coefficients, which are independent zero-mean circularly symmetric complex Gaussian variables, according to the path loss model using the linearized channel model of Fig. 1. In Fig. 3, we also provided the achievable rates for the non-causal case (corresponds to Theorem 2 with $\alpha = 0$) and the rate obtained with the ZF scheme, in which the secondary receiver considers the primary transmission as noise while the secondary transmitter uses ZF to cancel its own interference at the primary receiver. Note that, the rate obtained with the ZF scheme can be represented as

$$R_{\text{ZF}}^{(g)} = \max_{t} \left\{ p \log \left( 1 + \frac{P(1-t)}{p} \sum_{i=1}^{M} |h_{24,i}|^2 \right) + (1-p) \log \left( 1 + \frac{\frac{Pt}{(1-p)} \sum_{i=1}^{M-1} |\tilde{h}_{24,i}|^2}{1 + \frac{\beta P}{(1-p)M} \sum_{i=1}^{M} |h_{14,i}|^2} \right) \right\}.$$

The gain in degrees of freedom offered by the D-DPC-ZF scheme is evident in the figure, where the ZF scheme can only provide a constant SNR offset gain in the high SNR regime.

We finally note that, once we relax the first co-existence constraint, an upper bound on the sum-rate multiplexing gain follows by allowing full cooperation at both transmitters and receivers during the active primary periods. In such a scenario, multiplexing gain for the sum rate will be $\min(2M, 2)$ and hence an upper bound on the cognitive multiplexing gain can be obtained, where $r \leq 1$ due to second co-existence constraint. Here, although the time need to decode the primary message incurs a multiplexing gain loss of $(1-p)\frac{1}{M}$, the proposed scheme gets closer to be optimal in terms of multiplexing gain as the number of antennas at the transmitters increases. We finally note that the proposed scheme achieves the optimal multiplexing gain for the non-causal case.

## V. SISO-MIMO Channel

The next step is to analyze the scenario where the cognitive transmitter and receiver are equipped with additional antennas while maintaining a SISO primary channel. It is clear that, in this scenario, decoding at the cognitive transmitter will consume a prohibitive listening time resulting in a marginal performance gain in the high SNR regime. Interestingly, one can overcome this problem via two Zero-Forcing (ZF) filters. The first one is used by the cognitive transmitter to null-out the secondary message at the primary receiver. The second ZF filter is used by the cognitive receiver to null-out the primary message. The power levels used by the cognitive transmitter are then obtained from the spatial water-filling solution. The following result characterizes the achievable rate and optimal power allocation policy for the proposed scheme.

*Theorem 3:* An achievable rate of the SISO-MIMO cognitive channel using the proposed ZF scheme is given by:

$$R_{\text{ZF}}^{(g)} = \max_{t} \left\{ p \sum_{i=1}^{n_{min}} \log \left( 1 + P_i^* \lambda_i^2 \right) + (1-p) \sum_{i=1}^{\tilde{n}_{min}} \log \left( 1 + \tilde{P}_j^* \tilde{\lambda}_i^2 \right) \right\}, \quad (6)$$

where $P_i^*$s and $\tilde{P}_i^*$s are the water-filling power allocations: $P_i^* = \max\left(\mu - \frac{1}{\lambda_i^2}, 0\right), \tilde{P}_i^* = \max\left(\tilde{\mu} - \frac{1}{\tilde{\lambda}_i^2}, 0\right)$ with $\mu$ and $\tilde{\mu}$ are chosen to satisfy the total power constraints: $\sum_{i=1}^{n_{min}} P_i^* = \frac{P(1-t)}{p}, \sum_{i=1}^{\tilde{n}_{min}} \tilde{P}_i^* = \frac{Pt}{(1-p)}$, respectively. Here, $\lambda_i$s and $\tilde{\lambda}_i$s are the singular values of $\mathbf{H}_{24}$ and $\mathbf{Q}_r \mathbf{H}_{24} \mathbf{Q}_t$, respectively, where $\mathbf{Q}_t$ is a projection matrix at the cognitive transmitter, columns of which are orthogonal to $\mathbf{h}_{23} = [h_{23,1} \ldots h_{23,M}]$; $\mathbf{Q}_r$ is a projection matrix at the cognitive receiver, rows of which are orthogonal to $\mathbf{h}_{14} = [h_{14,1} \ldots h_{14,M}]^T$; and $\mathbf{H}_{24}^T$ is the matrix containing $h_{24,ij}$s.

The scheme achieves the optimal multiplexing gain, $r_{\text{ZF}}^{(g)} = pM + (1-p)(M-1)$, whereas the classical approach achieves a multiplexing gain of $r^{(c)} = pM$.

*Proof:* Please refer to Appendix IV. ∎

Here, we note that the cognitive zero forcing gain will disappear if $\tilde{\lambda}_i = 0 \ \forall i$. This situation corresponds to *a singular* channel in which the transmitter (or receiver) ZF cancels the secondary signal, as seen by the secondary receiver. Finally, Fig. 3 reports the performance gain of the proposed generalized SISO-MIMO cognitive channel, as compared with the classical approach with $M = 2$. To generate this figure, we used independent zero-mean circularly symmetric complex Gaussian channel coefficients, where each variances are set using the path loss model. The gain offered by the proposed approach is evident in the figure. Moreover, the gain in degrees of freedom, i.e., slope of the curve, is shown to approach $1 - p$ (the probability of the active state) as the SNR grows. Remarkable, imposing the coexistence constraints does not reduce the DoF compared to that of the MIMO interference channel with cooperation [9] for the SISO-MIMO setting.

## VI. CONCLUSION

In this work, we investigated the gain that can be leveraged from re-using the active areas in the frequency (time) spectrum of the cognitive radio channels. It was argued that this gain is limited, in the low and high SNR regimes, when the cognitive nodes are equipped with only single antennas. The limiting factor, in this scenario, was the need to decode and forward the primary signal at the cognitive transmitter in order to satisfy the *coexistence constraints*. This limitation was avoided in the MISO-MISO channel with the new D-DPC-ZF scheme where the decoding time, at the cognitive transmitter, does not become a limiting factor and forwarding the primary signal is avoided with a Zero-Forcing (ZF) approach. Then, we have proposed another cognitive transmission scheme for the SISO-MIMO channel which uses two ZF filters to cancel the secondary message at the primary receiver and vice-versa. Overall, our results shed light on the utility of frequency re-use in cognitive channel and illustrate the structure of efficient cognitive transmission schemes.

## APPENDIX I
## PROOF OF THEOREM 1

During state 1, cognitive transmitter transmits the cognitive message with an instantaneous rate of $R_1$.

$$R_1 = \log\left(1 + \frac{|h_{24}|^2 P(1-t)}{p}\right) \tag{7}$$

Then, after the first $N'$ channel use of state 2, if the information about primary message accumulated at the cognitive transmitter in $N'$ channel uses exceeds the total information at the primary receiver accumulated in $N$ channel uses, the secondary transmitter can decode the primary signal, which is a circularly symmetric complex Gaussian random variable $\sim C\mathcal{N}\left(0, \frac{\beta P}{(1-p)}\right)$. For a more detailed discussion, we refer the reader to [10]. Here we have

$$\sum_{n=1}^{N'-1} \log\left(1 + |h_{12}|^2 \frac{\beta P}{(1-p)}\right) \leq \sum_{n=1}^{N} \log\left(1 + |h_{13}|^2 \frac{\beta P}{(1-p)}\right) \leq \sum_{n=1}^{N'} \log\left(1 + |h_{12}|^2 \frac{\beta P}{(1-p)}\right).$$

In this inequality, dividing both sides by $N$, setting $\alpha = \lim_{N \to \infty} \frac{N'}{N}$, one can obtain the following

$$\alpha = \frac{\log\left(1 + \frac{|h_{13}|^2 \beta P}{(1-p)}\right)}{\log\left(1 + \frac{|h_{12}|^2 \beta P}{(1-p)}\right)} \tag{8}$$

and hence decoding is possible, if the channel gain between transmitters is better than the channel gain of the primary users, i.e., $\alpha < 1$.

During the remaining symbol durations of state 2, secondary transmitter will form the signal $x_2[n] = \sqrt{ut}d_2[n, x_1[n]] + \frac{h_{23}^*}{|h_{23}|}e^{j\angle h_{13}}\sqrt{\frac{(1-u)t}{\beta(1-\alpha)}}x_1[n]$, where $u \in [0, 1]$. Here, the first part of the signal is the dirty paper coded cognitive signal with known interference, i.e., $x_1[n]$ with a scale factor, and the second part is the primary signal, which is scaled according to power constraints and phase shifted to add-up coherently with the primary signal at the primary receiver. At this point, we remark that the resulting dirty-paper code is independent of the primary signal, and hence can be considered as noise at the primary receiver. See [6], [11] for details.

As the secondary transmitter uses the above signaling scheme, the instantaneous rates of the primary link ($R_p$) and the cognitive link ($R_c$) are as follows.

$$R_p = \log\left(1 + \frac{\left(|h_{13}|\sqrt{\beta} + |h_{23}|\sqrt{\frac{(1-u)t}{(1-\alpha)}}\right)^2}{\frac{|h_{23}|^2 ut}{(1-\alpha)} + \frac{(1-p)}{P}}\right) \quad (9)$$

$$R_c = \log\left(1 + \frac{|h_{24}|^2 P t u}{(1-\alpha)(1-p)}\right), \quad (10)$$

where $u$ is chosen such that $R_p = \log\left(1 + \frac{|h_{13}|^2 \beta P}{(1-p)}\right)$ in order to satisfy the second criteria of the *coexistence constraints* for $N' < n \leq N$. It follows that

$$u = 1 - \left(\frac{|h_{13}|\sqrt{\beta}\left(-\sqrt{(1-\alpha)}(1-p) \mp \sqrt{\delta}\right)}{|h_{23}|\sqrt{t}\left(1-p+|h_{13}|^2\beta P\right)}\right)^2 \quad (11)$$

$$\delta = (1-\alpha)(1-p)^2 + P|h_{23}|^2 t\left(1-p+|h_{13}|^2\beta P\right),$$

and $u \in [0,1]$. At this point, we remark that, since a continuous function assumes all of its intermediate values, a solution in $[0,1]$ exists. Finally, observing $R_2 = (1-\alpha)R_c$ and using $R_1$ and $R_2$ in (1) gives the achievable rate of the SISO cognitive link, i.e., for any $t \in [0,1]$,

$$R_{\text{DF-DPC}}^{(\text{g})}(t) = p\log\left(1 + \frac{|h_{24}|^2 P(1-t)}{p}\right) + (1-p)(1-\alpha)\log\left(1 + \frac{|h_{24}|^2 P t u}{(1-p)(1-\alpha)}\right)$$

is achievable, and hence, $R_{\text{DF-DPC}}^{(\text{g})} = \max_{t \in [0,1]} R_{\text{DF-DPC}}^{(\text{g})}(t)$ is achievable.

In the low SNR regime, for a given $t \in [0,1]$, we observe that

$$\lim_{P \to 0} \frac{R_{\text{DF-DPC}}^{(\text{g})}(t) - R^{(\text{c})}}{P} = \log(e)|h_{24}|^2(ut - t).$$

Now, the maximum in the low power regime can occur either at "$t = 0$" or at "$t \neq 0$ and $u = 1$", as $t, u \in [0,1]$. Here, both cases will result in $R_{\text{DF-DPC}}^{(\text{g})}(t) = R^{(\text{c})}$ and the latter case can not happen as $u = 1$ does not satisfy (11) as $P \neq 0$. Hence, we obtain

$$\lim_{P \to 0} \frac{R_{\text{DF-DPC}}^{(\text{g})} - R^{(\text{c})}}{P} = 0.$$

For the high power regime, considering the achievable rate $R_{\text{DF-DPC}}^{(\text{g})}(t)$ for a given $t \in [0,1]$,

$$\lim_{P \to \infty} R_{\text{DF-DPC}}^{(\text{g})}(t) - R^{(\text{c})} = \lim_{P \to \infty} p\log\left(1 + \frac{|h_{24}|^2 P(1-t)}{p}\right) + (1-p)R_2 - p\log\left(1 + \frac{|h_{24}|^2 P}{p}\right),$$

in which, using (5), we can show that $\lim_{P \to \infty} R_2 = 0$, for any $t \in [0,1]$. Hence, even with the maximization in considered in the expression of $R_{\text{DF-DPC}}^{(\text{g})}$, we can write

$$\lim_{P \to \infty} R_{\text{DF-DPC}}^{(\text{g})} - R^{(\text{c})} = 0.$$

## APPENDIX II
## PROOF OF COROLLARY 1

By setting $\alpha = 0$ and following an analysis similar to the one given in Appendix I, one can represent the achievable rate of the non-causal cognitive link as

$$R_{\text{F-DPC}}^{(\text{g,nc})} = \max_t \left\{ p \log\left(1 + \frac{|h_{24}|^2 P(1-t)}{p}\right) + (1-p)\log\left(1 + \frac{|h_{24}|^2 P t u^{(\text{nc})}}{(1-p)}\right) \right\}, \quad (12)$$

where using (5) with $\alpha = 0$, we have $u^{(\text{nc})} \in [0,1]$ satisfying

$$u^{(\text{nc})} = 1 - \left(\frac{|h_{13}|\sqrt{\beta}\left(-(1-p) \mp \sqrt{(1-p)^2 + P|h_{23}|^2 t \left(1 - p + |h_{13}|^2 \beta P\right)}\right)}{|h_{23}|\sqrt{t}\left(1 - p + |h_{13}|^2 \beta P\right)}\right)^2.$$

Now, we consider a power allocation $t \in [0,1]$, denote the corresponding achievable rate by $R_{\text{F-DPC}}^{(\text{g,nc})}(t)$, and after some algebra, we write

$$G(t) \triangleq \lim_{P \to \infty} R_{\text{F-DPC}}^{(\text{g,nc})}(t) - R^{(\text{c})} = p \log(1-t) + (1-p)\log\left(1 + \frac{|h_{24}|^2 \left(|h_{23}| t \mp 2|h_{13}|\sqrt{\beta t}\right)}{|h_{23}||h_{13}|^2 \beta}\right)$$

Here, it can be observed that, the optimal power allocation $t^*$ will approach to a fixed point in the interval $(0,1)$ as $P \to \infty$, which results in a constant rate gain (SNR offset gain) compared to the rate obtained with the classical approach. Therefore, we can write $\lim_{P \to \infty} R_{\text{F-DPC}}^{(\text{g,nc})} - R^{(\text{c})} = G$, where $G = \max_{t \in [0,1]} G(t)$.

## APPENDIX III
## PROOF OF THEOREM 2

During state 1, cognitive radio can use the capacity achieving transmission technique for MISO channels with transmit channel state information (CSI) and achieve the following rate.

$$R_1 = \log\left(1 + \frac{P(1-t)}{p}\sum_{i=1}^M |h_{24,i}|^2\right) \quad (13)$$

Then, cognitive radio enters into the listening phase, during which the received signals can be represented as follows:

$$\mathbf{y}_2[n] = \mathbf{H}_{12}\mathbf{x}_1[n] + \mathbf{z}_2[n] \quad (14)$$
$$y_3[n] = \mathbf{h}_{13}\mathbf{x}_1[n] + z_3[n], \quad (15)$$

where $\mathbf{H}_{12}{}^T$ is $M \times M$ matrix containing $h_{12,ij}$s, $\mathbf{h}_{13}$ is $1 \times M$ vector containing $h_{13,i}$s, $z_3 \sim \mathcal{CN}(0,1)$, $\mathbf{z}_2 \sim \mathcal{CN}(0, \mathbf{I}_M)$ and $\mathbf{x}_1 \sim \mathcal{CN}\left(0, \frac{\beta P}{(1-p)M}\mathbf{I}_M\right)$. Here, cognitive transmitter listens for the primary signal for $N'$ channel uses, where

$$\alpha = \lim_{N \to \infty} \frac{N'}{N} = \frac{\log\left(1 + \frac{P\beta}{(1-p)M}\mathbf{h}_{13}\mathbf{h}_{13}^H\right)}{\log \det\left[\mathbf{I}_M + \frac{P\beta}{(1-p)M}\mathbf{H}_{12}\mathbf{H}_{12}^H\right]} \quad (16)$$

After this listening period, if $\alpha < 1$, cognitive transmitter will send its vector channel input $\mathbf{x}_2[n] = [x_{2,1}[n], x_{2,2}[n], ..., x_{2,M}[n]]^T$ for $N' < n \leq N$. Here, as the primary message and hence the resulting interference at the secondary receiver is known, cognitive transmitter can construct its dirty-paper coded signal $d_2[n, x_1[n]]$. Now, to mitigate the interference at the primary receiver, cognitive transmitter forms an $M \times M - 1$ matrix $\mathbf{Q}_t$, which has orthonormal columns that are orthogonal to $\mathbf{h}_{23}$. Hence, the channel seen by the cognitive receiver becomes $\tilde{\mathbf{h}}_{24} = \mathbf{h}_{24}\mathbf{Q}_t$ and the channel input at the secondary transmitter can be formed as $\mathbf{x}_2[n] = \mathbf{Q}_t\tilde{\mathbf{x}}_2[n]$, where

$$\tilde{x}_{2,i}[n] = \frac{\tilde{h}_{24,i}^*}{\sqrt{\sum_{j=1}^{M-1}|\tilde{h}_{24,i}|^2}} d_2[n, x_1[n]], \text{ for } i = 1, \cdots, M.$$

During this transmission, the received signal at the cognitive receiver is

$$y_4[n] = \mathbf{h}_{14}\mathbf{x}_1[n] + \tilde{\mathbf{h}}_{24}\tilde{\mathbf{x}}_2[n] + z_4[n]. \tag{17}$$

Here, as $d_2[n, x_1[n]]$ is the dirty-paper coded cognitive signal constructed with the knowledge of the known interference, the achievable rate for the secondary user can be written as

$$R_2 = (1-\alpha)\log\left(1 + \sum_{i=1}^{M-1}|\tilde{h}_{24,i}|^2 \frac{Pt}{(1-\alpha)(1-p)}\right) \tag{18}$$

in the limit of large $N$. Finally, using $R_1$ and $R_2$ in (1) gives the achievable rate of the cognitive link for the MISO-MISO channel with the D-DPC-ZF scheme.

In addition, using the rate expressions above, we obtain

$$r_{\text{D-DPC-ZF}}^{(g)} = \lim_{P\to\infty}\frac{R_{\text{D-DPC-ZF}}^{(g)}}{\log(P)} = p + (1-p)\left(1 - \frac{1}{M}\right),$$

and

$$r^{(c)} = \lim_{P\to\infty}\frac{R^{(c)}}{\log(P)} = p.$$

## APPENDIX IV
## PROOF OF THEOREM 3

During state 1, cognitive link can utilize the channel only by itself as an $M \times M$ MIMO channel, where the channel coefficients are represented by $\mathbf{H}_{24}$, and can use the water-filling technique [12]. Here, denoting $n_{min} = M$ singular values of the matrix $\mathbf{H}_{24}$ with $\lambda_i$s, we will have

$$R_1 = \sum_{i=1}^{n_{min}}\log\left(1 + P_i^*\lambda_i^2\right), \tag{19}$$

where $P_i^*$s are the water-filling power allocations:

$$P_i^* = \max\left(\mu - \frac{1}{\lambda_i^2}, 0\right) \tag{20}$$

with $\mu$ is chosen to satisfy the total power constraint: $\sum_{i=1}^{n_{min}} P_i^* = \frac{P(1-t)}{p}$.

During the second state of the primary link, for every channel use $n \in \{1, 2, ..., N\}$, cognitive receiver will ZF the interference caused by primary transmission. To accomplish this task, cognitive receiver projects its received signal $\mathbf{y}_4[n]$ onto the subspace orthogonal to the one spanned by $\mathbf{h}_{14}$, which is $M \times 1$ vector containing $h_{14,i}$s. Since the projected space will have dimension $M-1$, projection matrix at the receiver, $\mathbf{Q}_r$, is an $M-1 \times M$ matrix, the rows of which are orthogonal to $\mathbf{h}_{14}$ and form an orthonormal basis of the subspace orthogonal to the one spanned by $\mathbf{h}_{14}$. We note that, since the rows are orthonormal, the processed vector, $\mathbf{Q}_r\mathbf{y}_4$, will still have i.i.d. unit-variance noises at each component. Similarly, cognitive transmitter using ZF strategy will cancel its interference at the primary receiver. This can be done by pre-multiplying the channel input by an $M \times M-1$ matrix, $\mathbf{Q}_t$, the columns of which are orthogonal to $\mathbf{h}_{23}$ and are chosen orthonormal.

Then, the cognitive link can utilize the channel as an $M - 1 \times M - 1$ MIMO channel, where the effective channel coefficients can be represented in the channel matrix $\tilde{\mathbf{H}}_{24} = \mathbf{Q}_r \mathbf{H}_{24} \mathbf{Q}_t$ [3], so that the received signal at the cognitive receiver is

$$\tilde{\mathbf{y}}_4[n] = \tilde{\mathbf{H}}_{24} \mathbf{x}_2[n] + \tilde{\mathbf{z}}_4[n] \tag{21}$$

where $\tilde{\mathbf{z}}_4 \sim C\mathcal{N}(0, \mathbf{I}_{M-1})$. The capacity of this cognitive link can be achieved by water-filling [12]. Denoting $\tilde{n}_{min}$ singular values of the matrix $\tilde{\mathbf{H}}_{24}$ with $\tilde{\lambda}_i$s, we will have

$$R_2 = \sum_{i=1}^{\tilde{n}_{min}} \log\left(1 + \tilde{P}_i^* \tilde{\lambda}_i^2\right), \tag{22}$$

where $\tilde{P}_i^*$s are the water-filling power allocations:

$$\tilde{P}_i^* = \max\left(\tilde{\mu} - \frac{1}{\tilde{\lambda}_i^2}, 0\right) \tag{23}$$

with $\tilde{\mu}$ is chosen to satisfy the total power constraint: $\sum_{i=1}^{\tilde{n}_{min}} \tilde{P}_i^* = \frac{Pt}{(1-p)}$. Then, using $R_1$ and $R_2$ in (1) gives the average rate of the cognitive link for the SISO-MIMO channel.

We next note that the proposed scheme achieves the optimal multiplexing gain. First, we compute the high SNR gain of the cognitive user using the ZF scheme. It is easy to see that cognitive user achieves a multiplexing gain of $M$ for each silent primary state and a multiplexing gain of $M - 1$ for each active primary state, where we assume that the channel matrices between users are not rank deficient. Hence, on the average, cognitive user achieves a multiplexing gain of $r_{ZF}^{(g)} = pM + (1-p)(M-1) = M - (1-p)$, whereas $r^{(c)} = pM$. We now state a result due to [13]. Denoting the number of antennas of Node i as $M_i$, for $i = 1, 2, 3, 4$, the total (sum-rate) DoF for the MIMO interference channel is given by

$$r^{\text{sum}}(M_1, M_2, M_3, M_4) \triangleq \min\{M_1 + M_2, M_3 + M_4, \max(M_1, M_4), \max(M_2, M_3)\}.$$

Recently, generalizing the single-antenna result of [8] to the multi-antenna setting, [9] has characterized the total (sum-rate) DoF for the MIMO interference channel with cooperation and it is shown that cooperation can not increase the total DoF. In other words, $r^{\text{sum}}$ given above is the total (sum-rate) DoF for the MIMO interference channel with cooperation. Returning to the problem formulation considered in this work, we see that due to the coexistence constraints, the multiplexing gain of the primary users is fixed at 1 during the active primary periods. Now, due to the outer bound provided by [9], the DoF that can be achieved by cognitive users during the active primary states is upper bounded by $r^{\text{sum}}(1, M, 1, M) - 1 = M - 1$, which is achieved by the proposed scheme. We therefore say that the cognitive users achieve the optimal multiplexing with the proposed scheme.

---

[3]We remark that the channel gains will still have Rayleigh distributions as each entry of $\mathbf{QH}$ has the same distribution as that of $\mathbf{H}$, where each entry of $\mathbf{H}$ is i.i.d. $\sim C\mathcal{N}(0, 1)$ and $\mathbf{Q}$ has orthonormal rows.

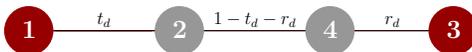

Fig. 1. Linear system model: Distance between transmitters and receivers are denoted as $t_d$ and $r_d$, respectively.

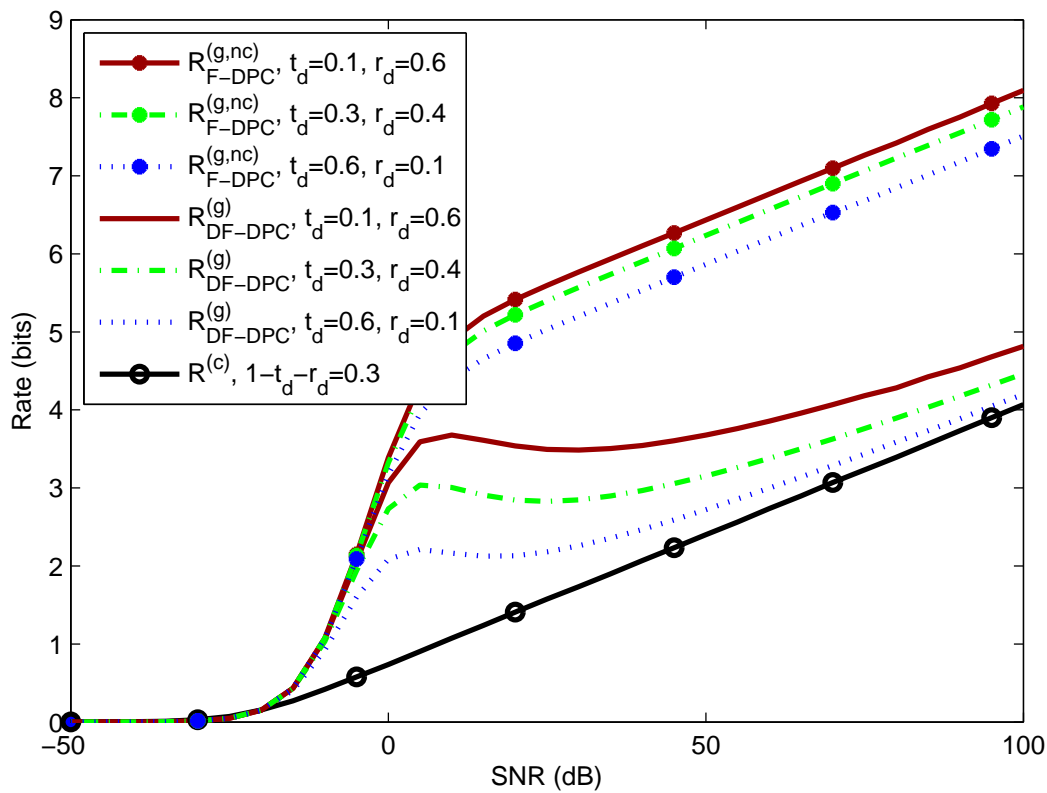

Fig. 2. Simulation results of the non-causal generalized cognitive radio (denoted with superscript (g,nc)), the causal generalized cognitive radio (denoted with superscript (g)), and the classical cognitive radio (denoted with superscript (c)) for the SISO-SISO channel with the linear system model given in Fig. 1 ($p = 0.1, \beta = 1$).

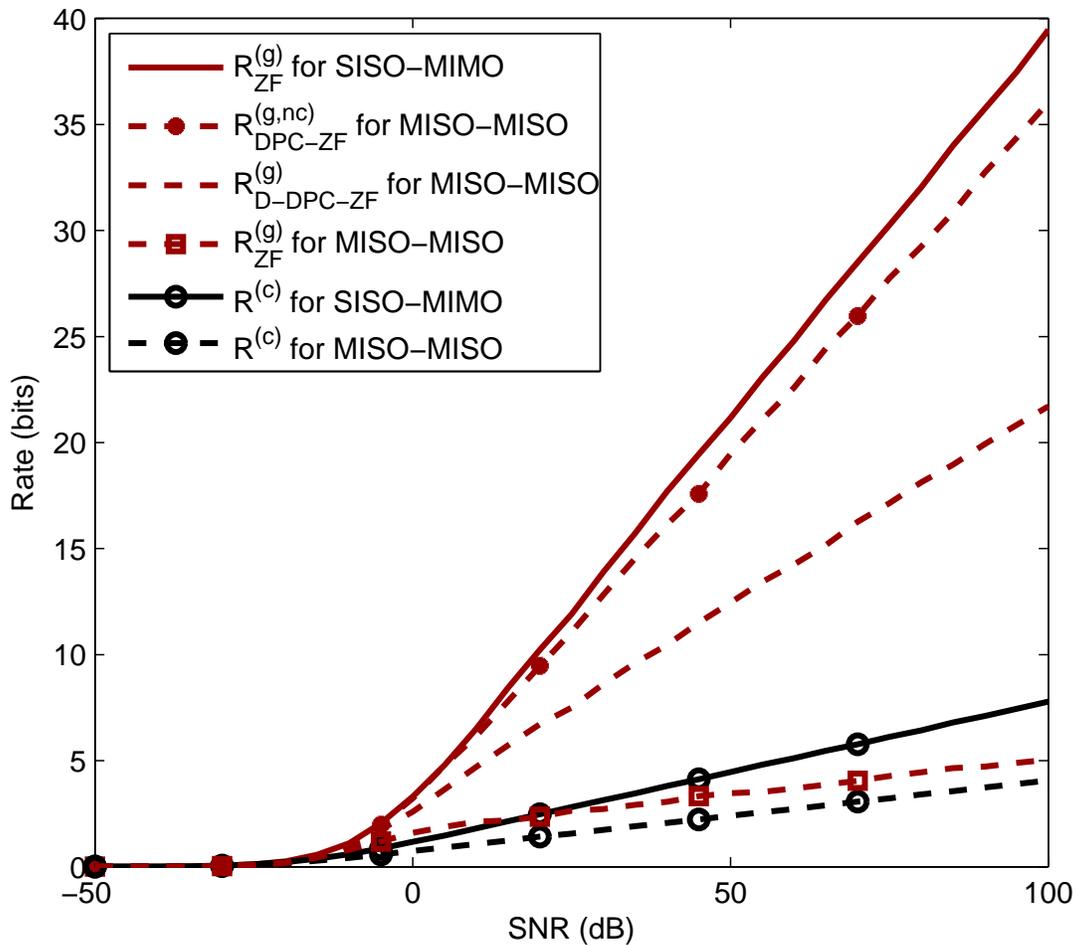

Fig. 3. Simulation results of the non-causal generalized cognitive radio (denoted with superscript (g,nc)), the causal generalized cognitive radio (denoted with superscript (g)), and the classical cognitive radio (denoted with superscript (c)) for the MISO-MISO channel (dashed lines) and the SISO-MIMO channel (solid lines) with Rayleigh fading channel coefficients. Variances of the channel gains are set according to the path loss model using the linearized system model of Fig. 1 ($p = 0.1, \beta = 1, M = 2, t_d = 0.1, r_d = 0.6$).